# Cancer cytoplasm segmentation in hyperspectral cell image with data augmentation

Rebeka Sultana[a], Hibiki Horibe[b], Tomoaki Murakami[b] and Ikuko Shimizu[c,*]

[a]*Institute of Global Innovation Research, Tokyo University of Agriculture and Technology, Tokyo, Japan*
[b]*Institute of Agriculture, Division of Animal Life Science, Tokyo University of Agriculture and Technology, Tokyo, Japan*
[c]*Institute of Engineering, Division of Advanced Information Technology & Computer Science, Tokyo University of Agriculture and Technology, Tokyo, Japan*

## ABSTRACT

Hematoxylin and Eosin (H&E)-stained images are commonly used to detect nuclear or cancerous regions in cells from images captured by a microscope. Identifying cancer cytoplasm is crucial for determining the type of cancer; hence, obtaining accurate cancer cytoplasm regions in cell images is important. While CMOS images often lack detailed information necessary for diagnosis, hyperspectral images provide more comprehensive cell information. Using a deep learning model, we propose a method for detecting cancer cell cytoplasm in hyperspectral images. Deep learning models require large datasets for learning; however, capturing a large number of hyperspectral images is difficult. Additionally, hyperspectral images frequently contain instrumental noise, depending on the characteristics of the imaging devices. We propose a data augmentation method to account for instrumental noise. CMOS images were used for data augmentation owing to their visual clarity, which facilitates manual annotation compared to original hyperspectral images. Experimental results demonstrate the effectiveness of the proposed data augmentation method both quantitatively and qualitatively.

## 1. Introduction

In the healthcare sector, digital pathology plays a crucial role in assisting pathologists in expediting disease diagnosis following a biopsy in the hospital's pathology laboratory (Shen (2022)). Hematoxylin and eosin (H&E) staining is a prevalent method in pathology for examining tissue morphology and structure. Hematoxylin stains cell nuclei blue or purple, accentuating nucleic acids, while eosin stains the extracellular matrix and cytoplasm pink, highlighting proteins and other cytoplasmic elements. H&E-stained images are commonly used to detect nuclear or cancerous regions in cells from images captured by a microscope.

To detect specific regions such as cells or nuclei from images, image segmentation has been extensively employed in medical imaging for various clinical applications, such as nuclear extraction for disease grading (Xing (2016)). Accurate segmentation of medical images is essential for precise diagnosis and comprehensive analysis. Although traditional segmentation methods have proven effective in numerous real-world scenarios, they often face challenges in detecting target regions in more complex cases. In this study, we focus on two aspects: data and methods.

First, we address the data aspect. While CMOS cameras are widely used for capturing pathology images, certain channels in CMOS images may lack sufficient information to accurately detect cancer. In such cases, hyperspectral imaging can provide significantly more detailed information about the pathology image (Ortega (2020)). Obtaining a hyperspectral image free from mechanical noise for pathological analysis is challenging. Although a stable microscope setup may yield a noise-free image, the variability in instrumental setups across different hospital environments means that noise frequently persists in hyperspectral images. Therefore, it is essential to consider instrumental noise in hyperspectral images for tasks such as segmentation.

Second, we discuss the method aspect. Identifying cancer cytoplasm in cell images is crucial for determining the type of cancer. For this purpose, the cancer cytoplasm area in cell images needs to be segmented. Traditional segmentation methods are not well-suited for identifying the cancer cytoplasm area owing to its complex visual features. In recent years, deep learning models have surpassed traditional methods in large-scale image classification tasks, marking the onset of a new era for their application in various tasks. Consequently, experiments have demonstrated the remarkable efficacy of deep learning models in computer vision tasks such as segmentation (Wang (2024)) and visual attention (Lai (2019)). However, a limitation of these models is that larger training datasets are required to improve their performance significantly, necessitating laborious and time-consuming data annotation for training purposes.

Acquiring large-scale hyperspectral pathology images poses a significant challenge. Additionally, training deep learning models in a supervised manner requires precise annotations, which are time-consuming to create. Since deep learning models necessitate extensive datasets to learn task-specific representations, pretraining/fine-tuning (Ghassemi (2019)) and data augmentation techniques (Perez (2017))

sultana@go.tuat.ac.jp (R. Sultana); s195215z@st.go.tuat.ac.jp (H. Horibe); mrkmt@cc.tuat.ac.jp (T. Murakami); ikuko@cc.tuat.ac.jp (I. Shimizu)

http://web.tuat.ac.jp/~tatlvt/ (T. Murakami); http://www.tuat.ac.jp/~ikuko/ (I. Shimizu)

ORCID(s): 0000-0001-7893-8048 (R. Sultana)

are frequently employed to address the limitations of small datasets.

To address the challenge of large-scale datasets in cell image segmentation, the TAJ-Net method generates fake samples and pseudo-labels for unlabeled data using a few-shot GAN model owing to the lack of high-quality training data (Zhang (2024)). Spectral images are used for segmentation, with adaptive joint training employed to eliminate noisy samples targeting cell type classification. In addition to GAN-based augmentation, test-time augmentation (TTA) is another technique used to handle the lack of massive data for large deep neural networks in cell image segmentation (Moshkov (2020)). It includes techniques like rotation, flipping, and merging; however, it excludes the addition of instrumental noise, which is crucial for our task. Well-known segmentation models such as U-Net (Ronneberger (2015)) and Mask-RNN (He (2017)) have also been used. In medical image segmentation, the modified U-Net model is favored because of its effectiveness in capturing important features ((Hatamizadeh (2022); Hatamizadeh (2021); Archit (2024); Guo (2022)). The goal is to achieve precise segmentation, where general augmentation is sometimes applied, as seen in (Hatamizadeh (2022)). Despite improvements in network architecture, a sufficient dataset is still required to train a model. As observed, most segmentation works in medical image segmentation involve data augmentation, whether using existing models or their modified versions. The approach applied depends on the dataset or the target task. Furthermore, no existing approach generates augmented datasets considering instrumental noise in hyperspectral image segmentation, particularly in cancer cytoplasm segmentation. Because instrumental noise is unavoidable when capturing hyperspectral images in many cases, it should be accounted for when handling noisy datasets. Additionally, real-world samples of cancer cytoplasm are limited. Generating fake images of different patterns using a GAN model outside the real-world samples is likely not a good choice because of the potential impact on future applications. Therefore, to the best of our knowledge, existing data augmentation techniques are not applicable for generating instrumental-noisy cancer cytoplasm datasets.

Instrumental noise is common in images captured by hyperspectral cameras base on the line push-broom method. For example, the publicly available near-infrared dataset of food items provided by HACARUS Inc., captured using the AVALDATA HS camera (AHS-U20MIR), contains line instrumental noise (Infrared). The Eva Japan NH7 captures images with significant line instrumental noise. It is caused by the sensitivity of the image sensor. Therefore, instrumental noise cannot be ignored in hyperspectral images captured by hyperspectral camera based on the line push-broom method. Since hyperspectral images provide various spectral information, instrumental noise must be considered when working with these kinds of images. In complex medical imaging tasks, such as segmentation or object detection using deep learning, manually annotating large datasets is challenging. Data augmentation that accounts for instrumental noise can be beneficial when working with hyperspectral images, as it reduces the annotation burden while maintaining superior performance. However, to the best of our knowledge, no existing work has considered instrumental noise for data augmentation in hyperspectral datasets.

This study proposes a method for using data augmentation techniques to create pseudo-hyperspectral images with instrumental noise from noise-free CMOS RGB pathology images of breast cancer. In the data augmentation process, CMOS images are first converted into grayscale, followed by brightness adjustment. Instrumental noise, both vertical and horizontal, is then added to the adjusted brightened image. Geometric transformation-based data augmentation techniques are applied to increase variation in the augmented datasets. Finally, the approach aims to train the well-known segmentation model U-Net (Ronneberger (2015)) with the augmented noisy images to segment the cancer cytoplasm area from hyperspectral pathology images. Experimental results confirm the effectiveness of the proposed method, both qualitatively and quantitatively.

The contributions of this work are summarized as follows:

1. This work presents, to the best of our knowledge, the first application of deep learning for segmenting cancer cytoplasm from hyperspectral pathology images.
2. For data augmentation, a CMOS RGB image is used to generate a pseudo-hyperspectral noisy image that closely resembles the original hyperspectral image.

## 2. Related work

This section provides an overview of existing histopathological image datasets used in medical image segmentation, commonly applied data augmentation techniques in medical image processing, and deep learning models used for segmentation tasks. It offers a comprehensive understanding of the current state of research in these areas. The details of each topic are explained in the following subsections.

### 2.1. Histopathological image dataset in medical image segmentation

In histopathological image analysis, various datasets are crucial for tasks such as classification, segmentation, regression, and registration. These datasets encompass different anatomical regions of the human body, including the breast, colon, liver, and bone. Table 1 provides a concise overview of H&E-stained image datasets primarily used for segmentation tasks, emphasizing the targeted anatomical structures and the presence of hyperspectral imagery of cancer cytoplasm.

The tabulated data underscores that the predominant focus of segmentation efforts is on nuclear and nuclei entities (Simon (2019); Graham (2021b)), aiming to alleviate the workload of pathologists in nuclear counting tasks. Other datasets are specifically employed for delineating tumor regions within tissue images (Kim (2021); Stadler (2021)), focusing on anatomical targets distinct from cancer cytoplasm. Notably, these datasets predominantly feature CMOS-acquired images.

**Table 1**
H&E-stained datasets for segmentation

| Organ | Dataset | Target object | Hyperspectral image |
|---|---|---|---|
| Cholangiocarcinoma | Sun (2022) | Tumor | ✓ |
| Lung | ACDC-LungHP Li (2020) | Lung carcinoma | |
| Liver | PAIP2019 Kim (2021) | Viable tumor | |
| Prostate | AGGC Huo (2022) | Gland | |
| Colorectal | GlaS Sirinukunwattana (2017) | Gland | |
| Colorectal adenocarcinoma | CoNSeP Simon (2019) | Nuclear | |
| Lymph node | CAMELYON16 Bejnordi (2017) | Metastases | |
| | CAMELYON17 Bándi (2019) | Metastases | |
| Skin | CATCH Wilm (2022) | Canine cutaneous tumor | |
| | Multi-Scanner SCC Wilm (2023) | Canine cutaneous tumor | |
| | CoCaHis Sitnik (2021) | Metastatic colon cancer | |
| Colon | CoNIC 2022 Graham (2021b) | Nuclei | |
| | CRAG Graham (2019) | Gland | |
| | Lizard Graham (2021a) | Nuclear | |
| | DigestPath2019 Jiahui (2019) | Cell | |
| | DROID Stadler (2021) | Tumor | ✗ |
| | Kumar (2017) | Nuclear | |
| | CryoNuSeg Mahbod (2021) | Nuclei | |
| | CPM-17 Vu (2019) | Nuclei | |
| Multiple | MoNuSAC Verma (2021) | Nuclei | |
| | MoNuSeg Kumar (2020) | Nuclei | |
| | PAIP2021 Ramin (2021) | Perineural invasion | |
| | SegPath Komura (2023) | Tissue | |
| | PanNuke Gamper (2020) | Nuclei | |
| | Janowczyk (2016) | Nuclei | |
| | TNBC Naylor (2019) | Nuclei | |
| Breast | BCSS Amgad (2019) | Tumor, stroma, inflammatory, necrosis | |
| | NuCLS Amgad (2022) | Nuclei | |
| | TIGER Shephard (2022) | Tumor, stroma | |
| | Gelasca (2008) | Cell | |
| | BACH Aresta (2019) | Benign, In situ carcinoma, invasive carcinoma | |
| | UCSB Gelasca (2009) | Nuclear | |
| | Our | Cancer cytoplasm | ✓ |

A H&E-stained dataset containing labels of tumor and non-tumor regions was used for automatic cholangiocarcinoma (Sun (2022)). Beyond conventional histopathological images, the Cx22 dataset (Ji (2023)) includes annotations for both cancer cytoplasm and nuclei, although its primary focus is on cell segmentation in cytological images. Contrarily, our dataset represents a unique effort, targeting the segmentation of cancer cytoplasm within hyperspectral images, with potential future applications in cancer-type identification.

## 2.2. Data augmentation in medical image analysis

A significant challenge in deep learning applications is the scarcity of training data, particularly in medical image analysis, where datasets are limited owing to patient consent constraints and the labor-intensive process of expert annotation.

To address this limitation, data augmentation techniques are crucial for enhancing the effectiveness of deep learning models trained on limited datasets. Commonly used augmentation methods include spatial transformations, adjustments in color and contrast, introduction of noise, application of deformations, and data mixing strategies (Cossio (2023); Garcea (2023); Goceri (2023)).

In medical image segmentation, specific augmentation techniques have been developed to address the unique characteristics of medical images. These include learned transformations (Zhao (2019)), mixup/mixmatch approaches (Eaton-Rosen (2019)), KeepMask/KeepMix techniques (Liu (2023)), and the application of AutoAugment methodologies, which incorporate standard augmentation strategies (Xu (2020)).

However, it is noteworthy that none of these studies have explicitly explored augmentation techniques for images with instrumental noise, a novel approach introduced in this paper.

### 2.3. Deep learning models in medical image segmentation

In 2015, the deep learning-based U-Net model significantly outperformed competitors in the cell tracking and segmentation competition at the IEEE International Symposium on Biomedical Imaging (Ronneberger (2015)). The U-Net model's sophisticated architecture has since led to its widespread adoption, with many subsequent segmentation models either directly based on U-Net or using U-Net architectures for various applications and dataset evaluations (Skourt (2018); Ziang (2020); Siddique (2021); Kim (2022)).

In addition to U-Net, other models such as DeepLabV3, DeepLabV3Plus, TransUNet, SegFormer, and DCAN are also employed for medical image segmentation tasks (Ji (2023); Chen (2017)). Given the proven effectiveness of U-Net-based models in segmentation, this study primarily used the U-Net model to evaluate the performance of the proposed data augmentation method.

## 3. Proposed method for Hyperspectral cell image segmentation

The method consists of two main components: the proposed data augmentation approach as the primary unit and the U-Net model for training as the secondary unit. The primary unit includes three key steps for data augmentation: color conversion and brightness adjustment, data augmentation through noise addition, and general data augmentation techniques. The secondary unit involves the use of a deep neural network (U-Net) trained on the augmented dataset for the segmentation task. The details of each unit are explained below.

### 3.1. Capturing technique of hyperspectral image and the reason of instrumental noise

The line push-broom method is one of the major spectroscopic imaging method. This imaging technique generates strips because a 2D image (XY) is divided into 1D (X) lines and spectroscopically separated. This process is repeated in the Y direction to create the 2D image (XY). The mechanism responsible for creating these lines is a slit. For instance, vertical stripes may appear if there is dust or other light-blocking material in part of the slit. Additionally, similar vertical stripes can occur due to variations in the sensitivity of the image sensor. Each manufacturer of hyperspectral cameras has different structures around the slit, but optically, they are generally similar. It has been confirmed that vertical stripes appear when capturing images by hyperspectral camera based on the line push-broom method. Dust is a mechanical cause, but since this is a line-by-line imaging method, post-processing may also influence the appearance of stripes. A 2D spectral image is generated from the spectral images of each line, but vertical stripes can arise during signal processing, such as when stitching these lines together.

Horizontal noise can also be present in images, potentially caused by fluctuations in the sensitivity of the image sensor or post-processing. There may also be additional bright and dark vertical stripes due to the sensor's sensitivity.

The H&E stained cell image captured by hyperspectral camera base on the line push-broom method contains various types of instrumental noise, as shown in Fig. 1. Some pixels may be shifted, and vertical or horizontal lines may be missing. Additionally, several lines may be filled with the same pixel values. As deep learning models require large-scale datasets, manually annotating noisy hyperspectral images is laborious and time-consuming. Therefore, data augmentation techniques can be applied while considering instrumental noise. In this case, CMOS images can be utilized. The characteristics of hyperspectral camera images differ from those of CMOS camera images, as lighting conditions vary significantly depending on the equipment used and its settings. Moreover, hyperspectral images can sometimes be extremely dark or bright, and the color of the light source can also vary considerably. CMOS images do not suffer from such issues. Therefore, in this work, CMOS images are used to generate pseudo-hyperspectral noisy images by converting the color space and adjusting brightness. Then, instrumental vertical and horizontal noise are added. Finally, to expand the dataset, general augmentation techniques are applied to the pseudo-hyperspectral noisy images.

### 3.2. Proposed data augmentation method

In the proposed method, the pseudo-noisy hyperspectral images are derived from CMOS images. A significant advantage of using CMOS images is their ability to generate multiple pseudo-noisy hyperspectral images from a single source image. The proposed data augmentation method is illustrated in Fig. 2, with each step detailed as follows:

*Color conversion and brightness adjustment:* Initially, the input CMOS image is converted into a grayscale format. Owing to excessive brightness in the grayscale image, gamma correction with a value of $\gamma$ is applied to adjust its luminance. This correction visually aligns the image with the hyperspectral image at specific bands that contain rich information for pathological diagnosis. Next, histogram equalization is applied to further enhance the image's visual fidelity. Finally, a single-color image filled uniformly with pixel values $C_1$ is added to produce the final spectral image.

*Data augmentation (Add noise):* Random horizontal and vertical noise lines are introduced into the spectral image, enabling the creation of $N$ pseudo hyperspectral images from a single spectral source. Vertical noise lines vary in number from $N_1$ to $N_2$, with a balanced distribution adjusted within a range of $\pm\sigma_1$ lines. Each noise line is represented by pixels with a value of $C_2$. Horizontal noise lines are positioned randomly within a region from $r_1$ to $r_2$, with variability of $\pm\sigma_2$ pixels. These horizontal lines contribute to information loss and pixel shifting within the image. Information loss is simulated by varying the slice height from $h_1$ to $h_2$ pixels, with a maximum loss of $m$ pixels. Pixel shifting involves displacing image information left or right by $d$ pixels. Consequently, three pseudo-noisy images are generated from a single CMOS image using this method.

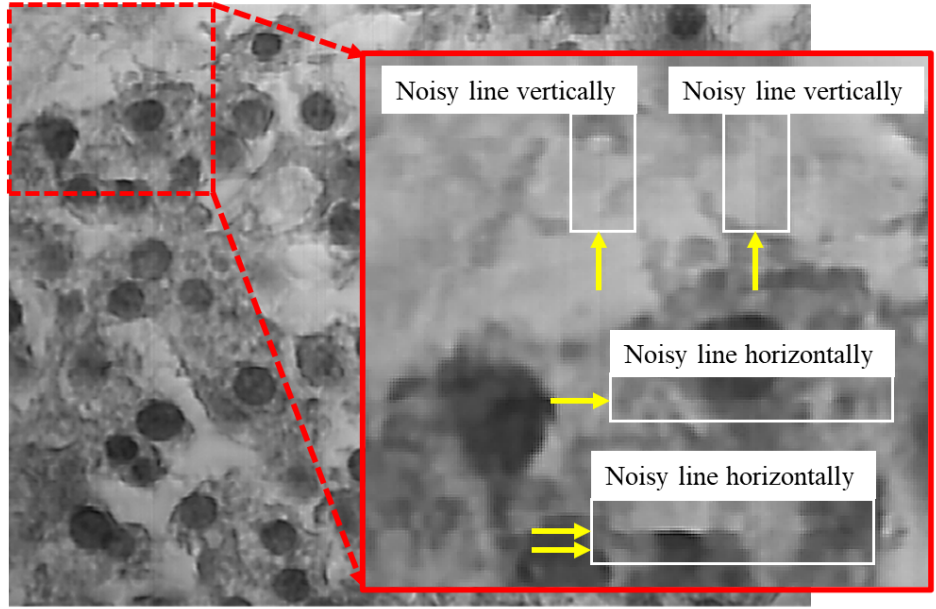

**Figure 1:** Instrumental noise in hyperspectral camera image

*Data augmentation (General):* To mitigate bias in image patterns caused by limited datasets, this study also employed general data augmentation techniques. The geometric transformations applied include cropping, horizontal flipping, translation, vertical flipping, and combined vertical-horizontal flipping. Cropping is performed randomly, with a maximum width of $c_w$ pixels and a maximum height of $c_h$ pixels. Each pseudo-noisy image undergoes $N$ different geometric transformations to produce a diverse set of augmented images.

### 3.3. U-net model architecture

The architecture of the U-net model employed in this study (Ronneberger (2015)) is depicted in Fig. 3. The feature size is expressed as channel × height × width. The one-channel hyperspectral input image undergoes convolution using a 3×3 kernel, followed by the ReLU activation function. This procedure is repeated, after which max pooling with a 2×2 kernel is applied to downsample the resulting feature. For an input image of size 1×480×640, the output feature size is 64×480×640 ($e_1$) before downsampling. The feature $e_1$ is then subjected to max pooling with a 2×2 kernel, followed by convolution, resulting in a feature size of 128×240×320 ($e_2$). This feature ($e_2$) undergoes further max pooling and convolution, yielding a feature size of 256×120×160 ($e_3$). Additional max pooling and convolution on $e_3$ produce a feature size of 512×60×80 ($e_4$). Further processing of $e_4$ results in a final feature size of 512×30×40 ($e_5$) in the contracting path. The feature $e_5$ is then upsampled using a 2×2 kernel, concatenated with $e_4$, and processed through convolution with ReLU activation to achieve a feature size of 512×60×80 ($d_4$), marking the initiation of the expansion path from the bottleneck. The feature $d_4$ is upsampled, concatenated with $e_3$, and convolved to produce a feature size of 256×120×160 ($d_3$). This pattern continues as $d_3$ is upsampled, concatenated with $e_2$, and convolved to achieve a feature size of 128×240×320 ($d_2$). Finally, $d_2$ is upsampled, concatenated with $e_1$, and convolved to obtain a feature size of 64×480×640 ($d_1$). A 1×1 convolution followed by sigmoid activation is then applied to $d_1$, resulting in the final output of size $N$×480×640, where $N$ represents the number of classes to be predicted, which is five in this study.

## 4. Experiments

The experimental section outlines the dataset acquisition procedure, including the environment system used, data annotation process, and experimental conditions during training. It also details the evaluation metrics employed to assess the model's performance. The Results and Discussion section presents the findings to validate the effectiveness of the proposed method. A detailed explanation of each subsection is provided below.

### 4.1. Dataset

H&E-stained dog mammary gland cells were used to prepare the dataset for this investigation. Imaging was conducted using both CMOS and hyperspectral cameras under a microscope. A total of 100 captured images are categorized into seven distinct types based on their visual patterns and color characteristics as shown in Fig. 4. Hyperspectral imaging involved acquiring 81 images for one H&E-stained sample across different wavelength bands, ranging from 400 to 800 nm at 5 nm intervals. Representative band images of different cells are shown in Fig. 5. The spectral band at 580 nm showed the brightest bands across all cell images. The brightness of the bands decreased gradually from 580 nm, leading to darker images, as shown in the spectral bands at 400 and 800 nm.

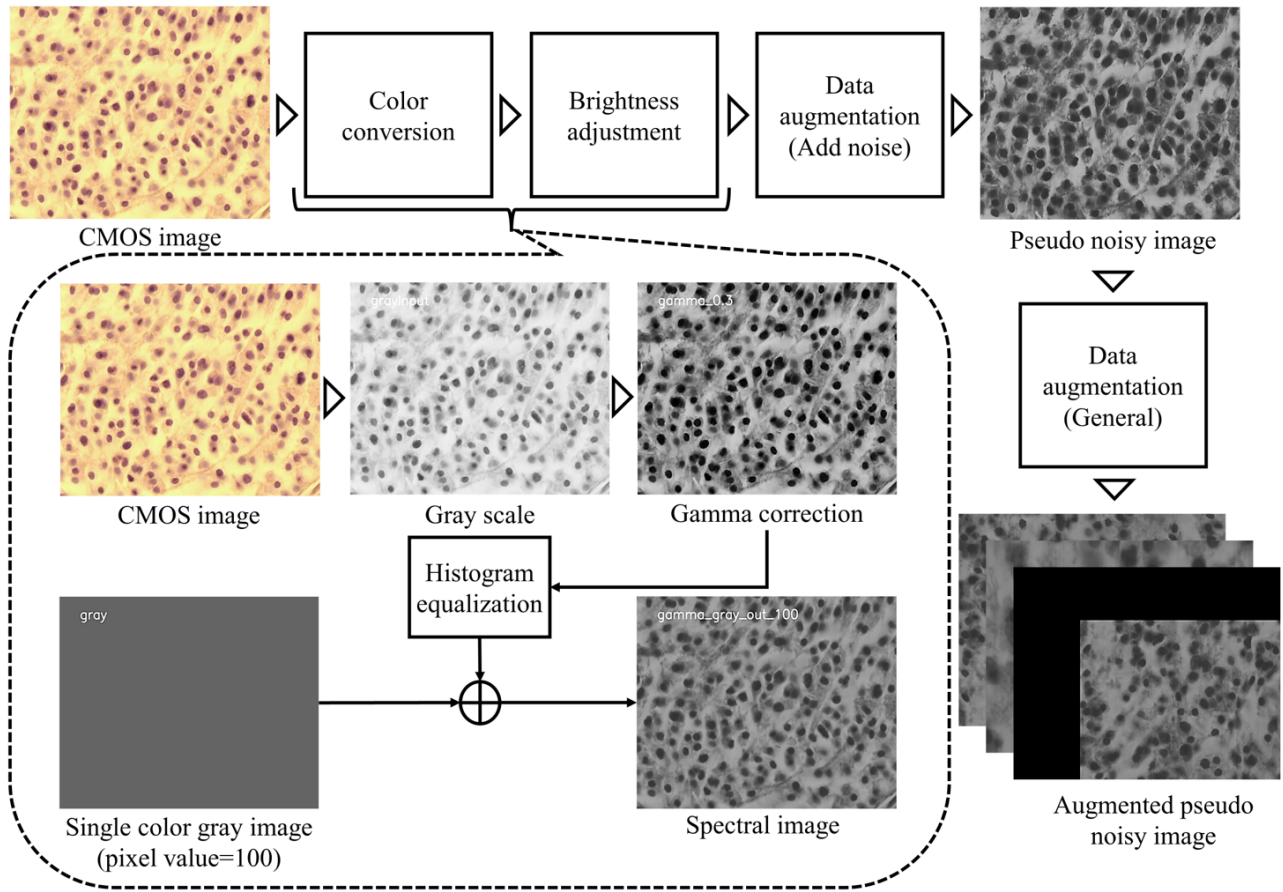

**Figure 2:** Proposed data augmentation method

The imaging system setup is shown in Fig. 6. The system consists of a microscope with a CMOS camera, hyperspectral camera placed on top of the microscope, and personal computer. The OLYMPUS BX51 microscope, equipped with an OLYMPUS UPlanFL objective lens and illuminated by a 100 W halogen lamp, was used to capture cellular images for this study. The WRAYCAM-VEX series CMOS camera was mounted on the microscope with its shutter speed set to automatic mode. The NH-1 series hyperspectral camera used a shutter speed of 16.64 ms per line during image capture. The personal computer used for capturing data was configured with an Intel® Core™ i7-9700 CPU @ 3.00 GHz, 16 GB of RAM. The operating system was Windows 11 Pro. Images from the hyperspectral camera exhibited instrumental noise, as shown in Fig. 1, characterized by both horizontal and vertical artifacts. Analysis of 100 hyperspectral images revealed approximately 24 noisy vertical lines, with distances between adjacent lines ranging from 2 to 76 pixels. Horizontal noise patterns were random, with 27 noisy lines observed across 24 hyperspectral images, spanning distances from 4 to 62 pixels. The 580 nm band was specifically selected for analysis owing to its optimal brightness among the band images examined.

## 4.2. Data annotation

Veterinary researchers in our group performed image annotations manually using the LabelMe software (labelme), and a veterinary professor validated the target areas to ensure their suitability for the experiment. A total of 58 images from both CMOS and hyperspectral types were annotated. In CMOS images, annotations included five categories: cancer cytoplasm, nuclear regions, red blood cells (RBCs), fibroblasts, and empty areas. Conversely, hyperspectral images were annotated only for cancer cytoplasm, nuclear regions, and empty areas. Sample annotated regions for both image types are illustrated in Fig. 7. Differences in the cellular field of view between CMOS and hyperspectral images were evident. Annotating hyperspectral data proved more time-consuming compared to annotating CMOS images owing to the textural complexities of cancer cytoplasm in grayscale images.

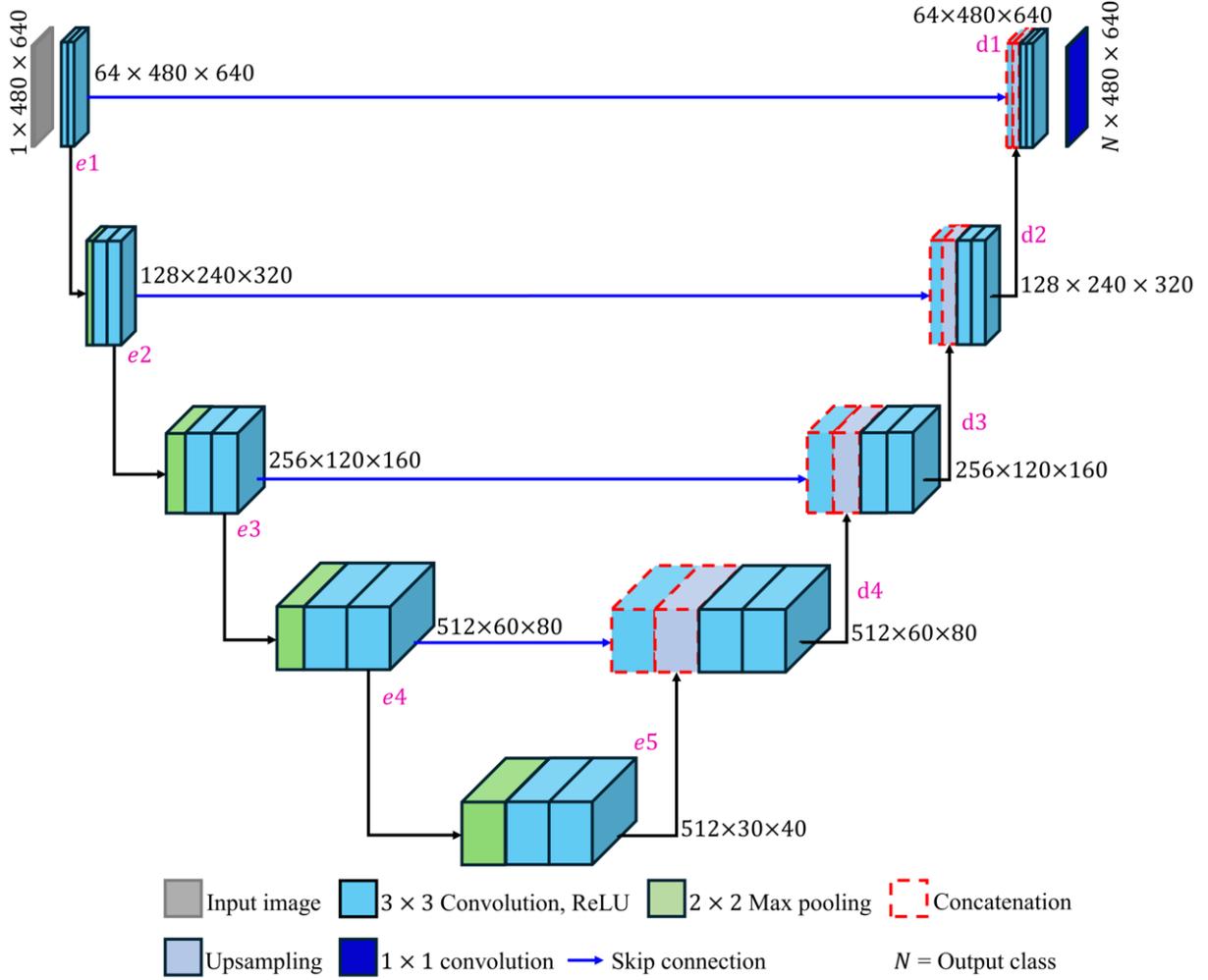

**Figure 3:** U-net model architecture

### 4.3. Experimental condition

For data augmentation, the parameters were as follows: $\gamma = 0.3$, $C_1 = 100$, $N_1 = 19$, $N_2 = 29$, $\sigma_1 = 5$, $C_2 = 128$, $r_1 = 26$, $r_2 = 32$, $\sigma_2 = 3$, $h_1 = 15$, $h_2 = 30$, $m = 2$, $d = 3$, $c_w = 800$, $c_s = 700$, and $N = 5$.

The experiment used 44 original CMOS images and 44 original hyperspectral images for training and augmentation. From the 44 CMOS images, a total of 792 pseudo-noisy images were generated. These pseudo-noisy images were combined with the original hyperspectral images, resulting in 836 images for training. Additionally, 56 original hyperspectral images of 7 types were used for evaluation.

The training parameters of U-Net and the experimental environment are presented in Table 2. The model was trained for 200 epochs with a batch size of 2. The learning rate was set at 0.0001, and RMSprop was used as the optimizer, with a weight decay of $1 \times 10^{-8}$ and a momentum of 0.9. Cross-entropy loss was used during model training because of the five segmentation classes. Training was conducted on a computer equipped with an Intel(R) Core(TM) i9-10900K CPU @ 3.70 GHz, 32 GB of RAM, and an NVIDIA GeForce RTX 3090 GPU with 24 GB of memory. The operating system was 64-bit Ubuntu 22.04.3 LTS.

### 4.4. Evaluation metrics

Two standard metrics, Intersection over Union (IoU) and the Dice coefficient, were employed to evaluate the performance of the trained model. IoU assesses the accuracy of the predicted segmentation map compared to the ground-truth map, while the Dice coefficient measures the similarity between the predicted and ground-truth segmentation maps. The mathematical expressions for these metrics are as follows:

$$\text{IoU} = \frac{\text{TP}}{\text{TP} + \text{FN} + \text{FP}}$$

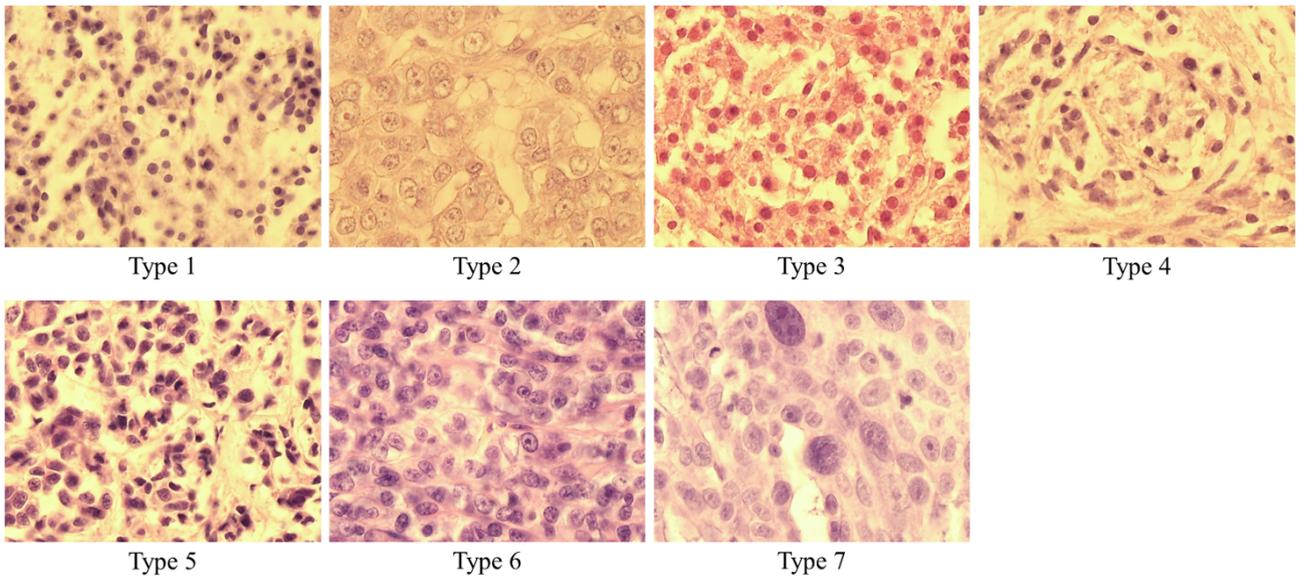

**Figure 4:** Example of different types of images

| Type | CMOS image | Spectral band: 400nm | Spectral band: 480nm | Spectral band: 580nm | Spectral band: 680nm | Spectral band: 780nm | Spectral band: 800nm |
|---|---|---|---|---|---|---|---|
| 1 | | | | | | | |
| 2 | | | | | | | |
| 3 | | | | | | | |
| 4 | | | | | | | |
| 5 | | | | | | | |
| 6 | | | | | | | |
| 7 | | | | | | | |

**Figure 5:** Example of hyperspectral images of different band of different cells

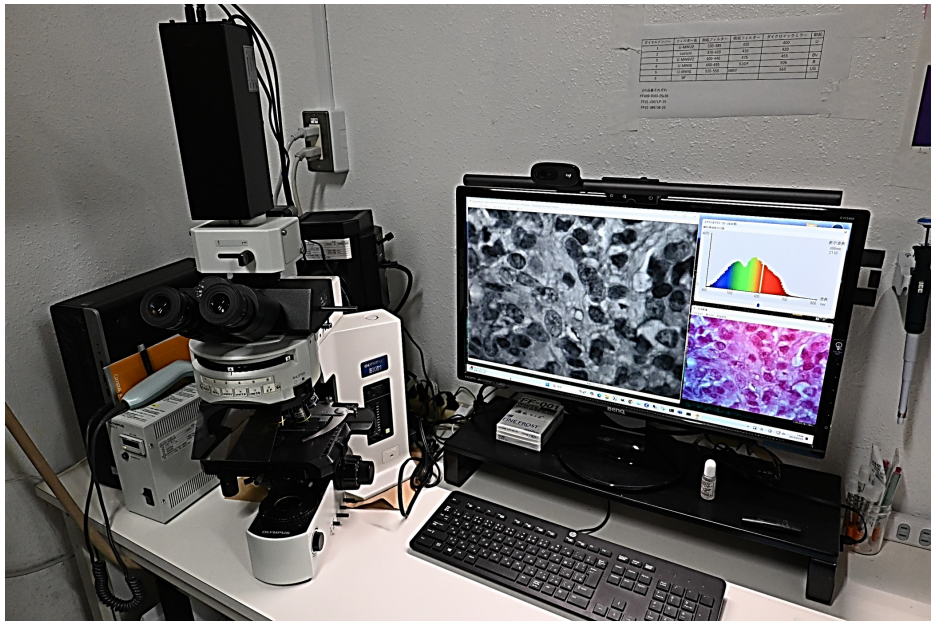

**Figure 6:** Imaging system

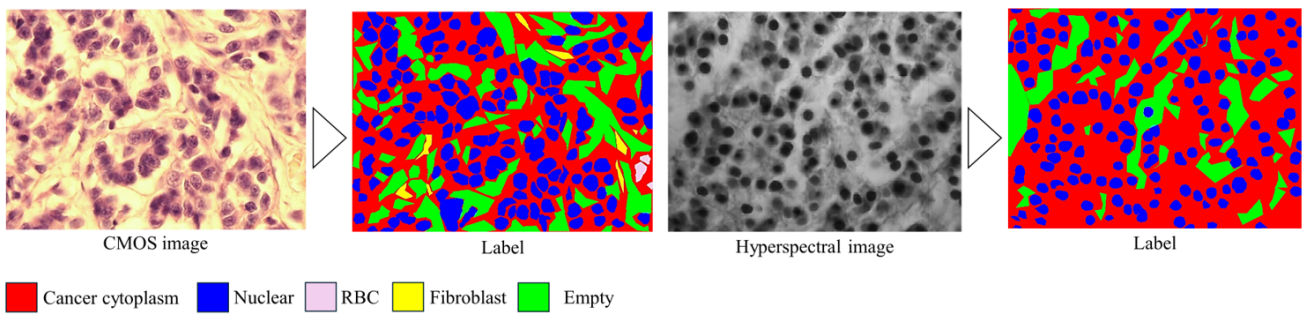

CMOS image    Label    Hyperspectral image    Label

🟥 Cancer cytoplasm   🟦 Nuclear   🟪 RBC   🟨 Fibroblast   🟩 Empty

**Figure 7:** Annotated data

where TP, FN, and FP indicate True Positive, False Negative, and False Positive, respectively.

$$\text{Dice Coefficient} = \frac{2 \times \text{TP}}{2 \times \text{TP} + \text{FP} + \text{FN}}$$

For both metrics, a score of 1.0 indicates that the predicted and ground-truth maps are identical, while a score of 0 signifies that the predicted and ground-truth maps are entirely dissimilar.

### 4.5. Result and Discussion

Figure 8 presents qualitative results evaluating predictions from different methods based on various patterns. Example (a) illustrates that the proposed method, highlighted by the yellow circle, showed less cancer cytoplasm compared to the comparison method and aligned more closely with the ground truth, which includes only the background. Example (b) reveals a false detection in the background region by the comparison method, whereas the proposed method's prediction more accurately matched the ground truth. In Example (c), the proposed method achieved more accurate predictions of the cancer cytoplasm area than the comparison method. Finally, Example (d) shows that the proposed method predicted a smaller background area within the cancer cytoplasm region compared to the comparison method. These qualitative observations validate the efficacy of the proposed method.

**Table 2**
Parameter and environment

| Trianing parameter | | Experimental environment | |
|---|---|---|---|
| Epoch | 200 | OS | Ubuntu 22.04.3 LTS |
| Batch size | 2 | OS type | 64-bit |
| Learning rate | 0.0001 | CPU | Intel(R) Core(TM) i9-10900K CPU @ 3.70GHz |
| Optimizer | RMSprop | CPU RAM | 32GB |
| Weight decay | $1 \times 10^{-8}$ | GPU | NVIDIA GeForce RTX 3090 |
| Momentum | 0.9 | GPU RAM | 24GB |
| Loss | Cross entropy | - | - |

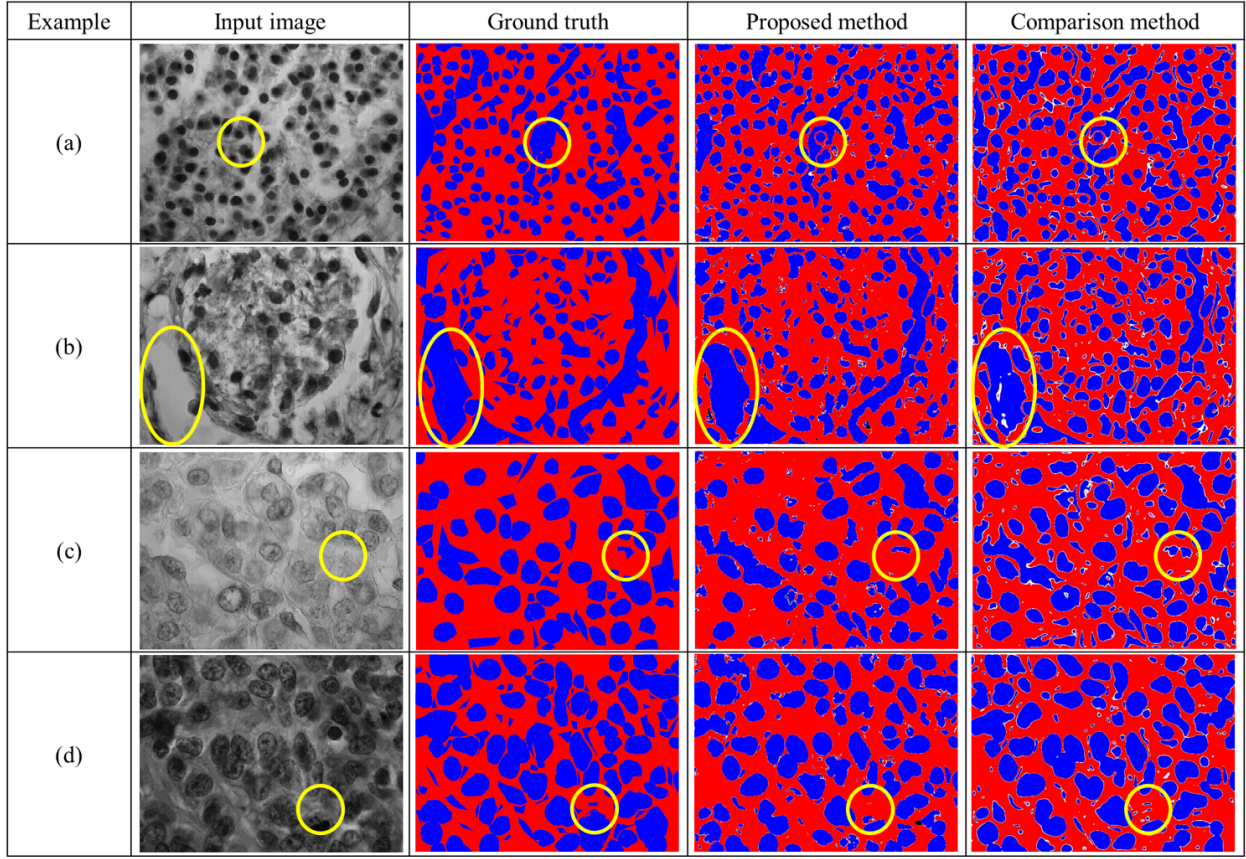

| Example | Input image | Ground truth | Proposed method | Comparison method |
|---------|-------------|--------------|-----------------|-------------------|
| (a) | | | | |
| (b) | | | | |
| (c) | | | | |
| (d) | | | | |

**Figure 8:** Qualitative results. The red area represents the cancer cytoplasm, while blue indicates other areas. White and dark brown highlight areas with false detections, where the predicted cancer cytoplasm overlaps with other areas. The comparison method was trained using only small-scale original hyperspectral noisy images. The yellow circle indicates the area of focus.

**Table 3**
Quantitative results (Data Augmentation). (↑) indicates that higher the score, the better is the model performance is.

| Method | Number of train images | Image type | IoU (↑) | Dice coefficient (↑) |
|--------|------------------------|------------|---------|----------------------|
| 1 (comparison) | 44 | Original noisy hyperspectral image | 0.7255 | 0.8392 |
| 2 (proposed) | 836 | Original and pseudo noisy hyperspectral image | 0.7362 | 0.8466 |

Table 3 lists the quantitative results. Trained on the original hyperspectral images, U-Net, referred to as Method 1, achieved an IoU of 0.7255 and a Dice coefficient of 0.8392 when tested with original noisy hyperspectral images. Introducing pseudo-noisy hyperspectral images alongside the originals for training improved the IoU and Dice coefficient by 1.5% and 0.9%, respectively, highlighting the effectiveness of the proposed data augmentation method.

The effectiveness of varying the number of augmented images was evaluated through ablation studies, as detailed in Table 4. Method 1, which generated training images from CMOS images, achieved an IoU of 0.6535 and a Dice coefficient of 0.7857. Method 2 showed that increasing the quantity of pseudo-noisy images improved the model's performance by 3.4% and 2.4% in terms of IoU and the Dice coefficient, respectively, highlighting the benefits of noise addition through data augmentation. Method 3 expanded on this by incorporating geometrically transformed augmented datasets, which increased the training image count and boosted model performance by 4.5% and 2.7% in terms of IoU and the Dice coefficient, respectively. Notably, Method 4 (proposed) further improved segmentation model performance by incorporating the original noisy hyperspectral image along with 792 additional images, resulting in gains of 4.3% and 2.5% in terms of IoU and the Dice coefficient, respectively, compared to Method 3. These ablation studies conclusively validate the efficacy of data augmentation techniques involving instrumental noise and geometric transformations in cancer cytoplasm segmentation.

**Table 4**
Ablation analysis on some augmented images in training. (↑) indicates that the higher the score, the better is the model performance is.

| Method | Original hyperspectral image | Pseudo hyperspectral image | IoU (↑) | Dice coefficient (↑) |
|---|---|---|---|---|
| 1 | | 44 | 0.6535 | 0.7857 |
| 2 | ✗ | 132 | 0.6758 | 0.8042 |
| 3 | | 792 | 0.7059 | 0.8258 |
| 4 (proposed) | 44 | 792 | 0.7362 | 0.8466 |

**Table 5**
Effectiveness of instrumental noise using pseudo images

| Method | Noisy | Color adjusted | General augmentation | IoU (↑) | Dice coefficient (↑) |
|---|---|---|---|---|---|
| 1 | ✓ | | ✗ | 0.6535 | 0.7857 |
| 2 | | | | 0.6485 | 0.7833 |
| 3 | | | Cropping | 0.6628 | 0.7943 |
| 4 | ✗ | ✓ | Horizontal flipping | 0.6423 | 0.7789 |
| 5 | | | Translation | 0.6249 | 0.7658 |
| 6 | | | Vertical flipping | 0.6449 | 0.7813 |
| 7 | | | Vertical and horizontal flipping | 0.4012 | 0.5135 |

To demonstrate the effectiveness of the pseudo-noisy images, general augmentation techniques were applied to the color-adjusted images from the CMOS sensor. Table 5 lists the impact of noisy images on cancer cytoplasm prediction. Forty-four images were used for training, while the test set remained the same with 56 original hyperspectral noisy images. When the images contained only pseudo noise as the augmentation technique, the IoU and Dice coefficients were the highest compared to images without noise except Method 2. General augmentation techniques, such as horizontal flipping, vertical flipping, translation, and combinations of vertical and horizontal flipping, were applied to images without noise. The evaluation scores did not surpass those of the model trained on noisy images. On the other hand, Method 2, trained with only cropping augmentation, outperforms Method 1 because the close-up segmentation views in the training dataset likely help the model learn standout classes, such as the nuclear area, more effectively. As Method 3 to 7 use images without altering the position of segmentation classes, this result suggests that pseudo-noisy training images are crucial for segmenting cancer cytoplasm in the dataset.

In addition to the proposed pseudo-noise, five general augmentation techniques were included to increase the number of training images. The effectiveness of these augmentation techniques was validated by training the models on pseudo-noisy images with these general augmentation methods. First, each augmentation technique was applied individually to the pseudo-noisy images to identify the most and least effective techniques, as summarized in Table 6. The number of pseudo-noisy images was 132. Subsequently, each augmentation technique was applied sequentially to obtain 264 training images. Cropping was the most effective augmentation technique, resulting in higher IoU and Dice coefficients compared with other techniques. Horizontal flipping was the least effective augmentation technique. These evaluation scores confirm that proper augmentation can enhance prediction scores.

**Table 6**
Effectiveness of general augmentation technique using pseudo images (single)

| Method | General augmentation | Number of image | IoU (↑) | Dice coefficient (↑) |
|---|---|---|---|---|
| 1 | ✗ | 132 | 0.6758 | 0.8042 |
| 2 | Cropping | | 0.6972 | 0.8199 |
| 3 | Horizontal flipping | | 0.6473 | 0.7833 |
| 4 | Translation | 264 | 0.6565 | 0.7901 |
| 5 | Vertical flipping | | 0.6667 | 0.7977 |
| 6 | Vertical and horizontal flipping | | 0.6572 | 0.7907 |

**Table 7**
Effectiveness of general augmentation technique using pseudo images (combination)

| Method | General augmentation | Number of images | IoU (↑) | Dice coefficient (↑) |
|---|---|---|---|---|
| 1 | Cropping | 264 | 0.6972 | 0.8199 |
| 2 | Cropping, vertical flipping | 396 | 0.7000 | 0.8219 |
| 3 | Cropping, vertical flipping, translation | 528 | 0.7000 | 0.8217 |
| 4 | Cropping, vertical flipping, translation, vertical and horizontal flipping | 660 | 0.6899 | 0.8146 |
| 5 | Cropping, vertical flipping, translation, vertical and horizontal flipping, horizontal flipping | 792 | 0.7059 | 0.8258 |

**Table 8**
Evaluation scores of each type of images

| Type | Proposed | Comparison | Proposed | Comparison |
|------|----------|------------|----------|------------|
|      | IoU      |            | Dice Coefficient |    |
| 1 | 0.7477 | 0.7244 | 0.8548 | 0.8391 |
| 2 | 0.7021 | 0.6930 | 0.8237 | 0.8173 |
| 3 | 0.7032 | 0.6806 | 0.8218 | 0.8062 |
| 4 | 0.7281 | 0.6812 | 0.8414 | 0.8083 |
| 5 | 0.7342 | 0.7484 | 0.8460 | 0.8558 |
| 6 | 0.7456 | 0.7470 | 0.8536 | 0.8546 |
| 7 | 0.7652 | 0.7624 | 0.8662 | 0.8645 |

**Table 9**
Evaluation scores of different number of test images

| Number | Proposed | Comparison | Proposed | Comparison |
|--------|----------|------------|----------|------------|
|        | IoU      |            | Dice Coeffient |    |
| 14 | 0.7657 | 0.7169 | 0.8661 | 0.8326 |
| 21 | 0.7420 | 0.7134 | 0.8501 | 0.8306 |
| 28 | 0.7378 | 0.7176 | 0.8474 | 0.8336 |
| 35 | 0.7365 | 0.7178 | 0.8466 | 0.8337 |

Next, using cropping as the base general augmentation technique, the best two, three, and four augmentation techniques, as well as all five, were applied to assess the effectiveness of general augmentation in the proposed method, as presented in Table 7. Cropping and vertical flipping improved the IoU and Dice coefficients by 0.4% and 0.2%, respectively, compared to using cropping alone. Adding translation with cropping and vertical flipping slightly increased the evaluation score of Dice coefficient while IoU remains same. However, adding both vertical and horizontal flipping with the previous three augmentation techniques led to a decrease in the evaluation scores. Finally, when all five general augmentation techniques were applied to generate the training dataset, the model achieved the highest IoU and Dice coefficients among all models, as summarized in Table 7. Therefore, in addition to pseudo-noisy augmentation, general augmentation plays a vital role in increasing the number of training images for more precise segmentation. As the proposed method targets an instrumentally noisy hyperspectral dataset to achieve precise segmentation, other tasks, such as object detection, may also benefit from it when applied to similar types of hyperspectral datasets in the future.

The above experiments highlight the effectiveness of the proposed data augmentation. Since there are only seven types of images in the dataset, the performance of each type on the test set is also analyzed. The results are shown in Table 8. It can be seen that the proposed method outperforms the comparison method except for Types 5 and 6. It is assumed that this is because the model is biased toward certain types. Therefore, the mean image of each test set, along with histogram analysis, is calculated, as shown in Fig. 9. The mean images of Types 5 and 6 are much darker than those of other types. Moreover, along with these two types, Type 7 contains more congested nuclei than other types of images. However, from the figure, it can be seen that Type 7 is lighter than Types 5 and 6. Although the comparison results in Table 8 show that Types 5 and 6 perform better than the proposed method, the results for Types 1, 2, 3, and 4 indicate relatively lower scores for the comparison method than for the proposed method. The comparison method achieves scores close to the proposed method for Type 7. This suggests that when the model is trained on a small number of original images, it learns to recognize darker image types well. On the other hand, since the training set in the proposed method contains a large number of pseudo-images created by adding a single-color fixed-value gray image, the model appears to minimize bias toward brightness as well as structural patterns. Based on these observations, it can be suggested that including more darker pseudo-images could improve the performance of the proposed method.

Since the number of evaluation images is only 56, and the different types of images are not evenly distributed, the number of each type was increased from 2 to 5, resulting in 14, 21, 28, and 35 images, respectively, to analyze how the model performs with a small number of evaluation samples. The maximum number was set to 5 because each type contains at least 5 images. The quantitative results are shown in Table 9. From the table, it can be seen that when evenly distributed images are used to prevent bias, the overall scores for the proposed method are better than those of the comparison method. This indicates the potential of the proposed method, even with a small evaluation dataset.

## 5. Limitations

The main limitation of the proposed method is its reliance on datasets containing instrumental noise for optimal performance. The dataset must be expanded to include more diverse patterns to enhance model robustness and support the development of advanced deep learning models with large parameters capable of producing precise segmentation maps. Incorporating additional noisy samples derived from original cell samples could improve the representativeness and utility of the dataset. Although obtaining sufficient data from real samples is challenging, it may be feasible to gradually collect more data over time, ultimately enabling more accurate segmentation maps for cancer cytoplasm detection.

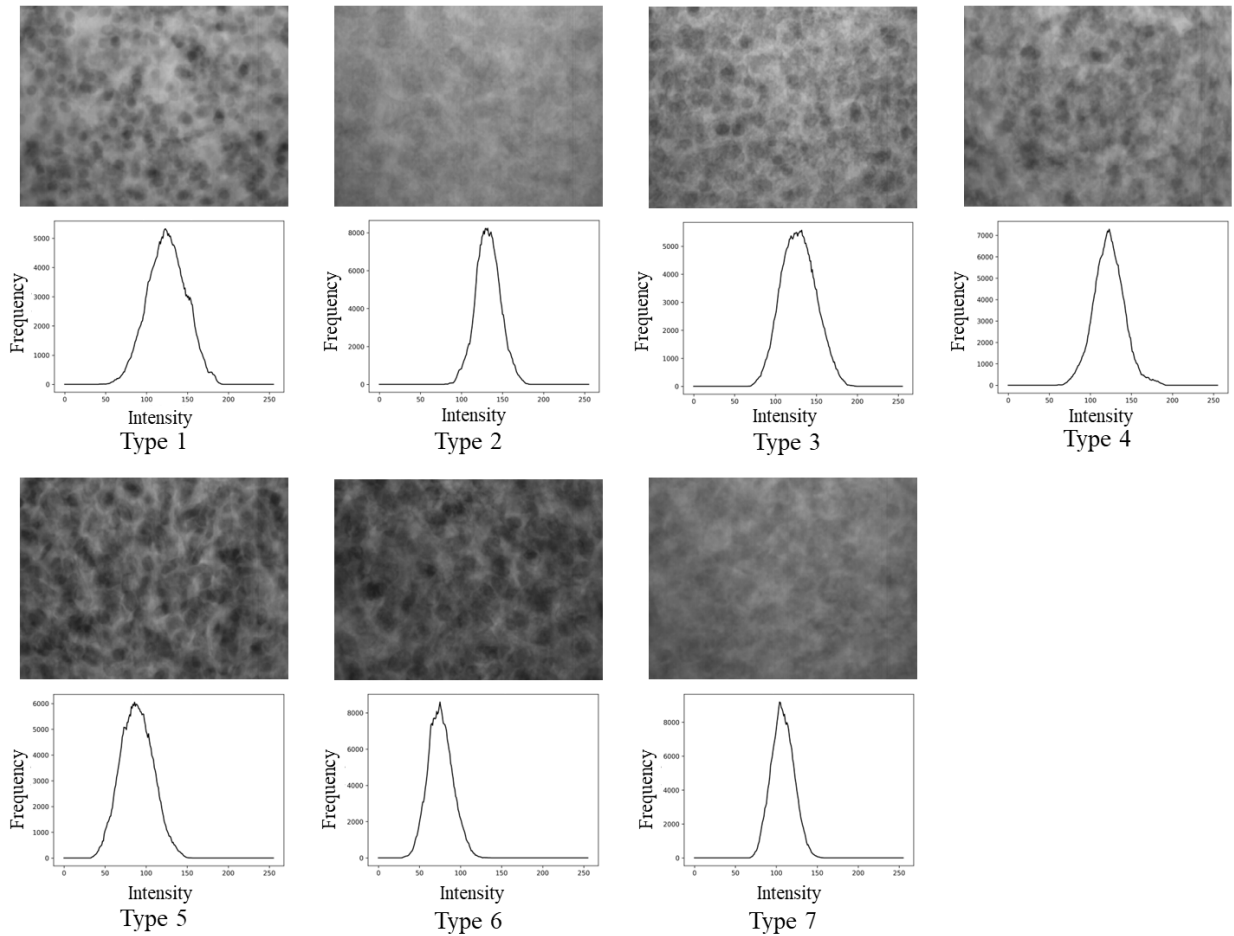

**Figure 9:** Mean image of each type of test images

## 6. Conclusion

The identification of cancer cytoplasm in H&E-stained cell images is crucial for determining cancer type. Hyperspectral imaging provides more detailed data compared to CMOS imaging. However, instrumental noise caused by microscope placement in laboratory environments can significantly impact results. To the best of our knowledge, no research has specifically focused on segmenting cancer cytoplasm from cell images using hyperspectral imaging while accounting for instrumental noise. Given the need for large-scale datasets to train effective deep learning models, this study proposed a data augmentation approach that integrates instrumental noise into hyperspectral images to improve cancer cytoplasm segmentation using a U-net model. The augmentation process involves converting CMOS images into grayscale, adjusting brightness, and incorporating both instrumental noise and geometric transformations. Experimental results validate the efficacy of the proposed method both qualitatively and quantitatively.

## Acknowledgment

This work was supported by the TUAT Fusion Research Support System TAMAGO.

## CRediT authorship contribution statement

**Rebeka Sultana:** Methodology, Investigation, Software, Visualization, Writing draft. **Hibiki Horibe:** Data curation. **Tomoaki Murakami:** Funding acquisition, Resources, Validation. **Ikuko Shimizu:** Project administration, Supervision, Review & editing.